\documentclass[twocolumn,prl,groupedaddress,10pt]{revtex4}

\usepackage{graphicx}
\usepackage[usenames,dvipsnames]{color}
\usepackage{amssymb}
\usepackage{amsmath}
\usepackage{empheq}
\usepackage{float}

\begin{document}

\title{Friction as Contrast Mechanism in Heterodyne Force Microscopy}

\author{G.J. Verbiest} \email[]{Verbiest@physics.leidenuniv.nl}
\affiliation{Huygens-Kamerlingh Onnes Laboratory, Leiden University, Niels Bohrweg 2, 2333 CA Leiden, The Netherlands\\Current address: JARA-FIT and 2nd Institute of Physics, RWTH Aachen University, 52074 Aachen, Germany}
\author{T.H. Oosterkamp}
\affiliation{Huygens-Kamerlingh Onnes Laboratory, Leiden University, Niels Bohrweg 2, 2333 CA Leiden, The Netherlands}
\author{M.J. Rost} \email[]{Rost@physics.leidenuniv.nl}
\affiliation{Huygens-Kamerlingh Onnes Laboratory, Leiden University, Niels Bohrweg 2, 2333 CA Leiden, The Netherlands}

\date{\today}

\begin{abstract}
The nondestructive imaging of subsurface structures on the nanometer scale has been a long-standing desire in both science and industry. A few impressive images were published so far that demonstrate the general feasibility by combining ultrasound with an Atomic Force Microscope. From different excitation schemes, Heterodyne Force Microscopy seems to be the most promising candidate delivering the highest contrast and resolution. However, the physical contrast mechanism is unknown, thereby preventing any quantitative analysis of samples. Here we show that friction at material boundaries within the sample is responsible for the contrast formation. This result is obtained by performing a full quantitative analysis, in which we compare our experimentally observed contrasts with simulations and calculations. Surprisingly, we can rule out all other generally believed responsible mechanisms, like Rayleigh scattering, sample (visco)elasticity, damping of the ultrasonic tip motion, and ultrasound attenuation. Our analytical description paves the way for quantitative SubSurface-AFM imaging.
\end{abstract}


\maketitle

\section{Introduction}

Many fields of research are in need of a nondestructive way of imaging nanometer-sized subsurface features. To this end, ultrasound was combined with Atomic Force Microscopy to invent (Waveguide-) Ultrasound Force Microscopy \cite{Kolosov,Yam} and Heterodyne Force Microscopy (HFM) \cite{Cub3}. HFM makes use of two ultrasound waves at slightly different frequencies, one of which is sent through the sample and the other through the cantilever. The mixed, heterodyne signal (amplitude and phase) at their frequency difference contains possible subsurface information at an experimentally accessible frequency \cite{Garcia3}.

Using HFM, subsurface images with remarkable contrast and resolution have been reported \cite{Striegler,Rabe2,Hu,Vitry,Dra1,Dra2,Cant1,Tet1,Tet2,Tet4,Tet5,Tet6,Cub2,Cub3,Yam2}, like the detection of 17.5 nm large gold nanoparticles buried at a depth of 500 nm in a polymer \cite{Dra1}. Surprisingly, the generated contrast clearly exceeds the background variations in these images, although the size of the nanoparticles is only a fraction of the sample thickness, and the lateral fingerprint {\it on} the surface (resolution) is equal to the diameter of the nanoparticles. Both observations are hard to understand, if one considers the wavelengths of the ultrasonic excitations, which is in the order of mm's and therefore much larger than both the size of the nanoparticles (nm's) and their depth below the surface (up to $\mu$m's). Unfortunately, none of the published HFM experiments provides quantitative information on the measured amplitude and phase range, on the applied contact force during the measurement, and on the precise excitation scheme in combination with the resonance frequencies of the cantilever.

To pave the way for quantitative subsurface measurements, it is of crucial importance to understand the physical contrast formation mechanism \cite{Garcia}. This requires a detailed, quantitative understanding of the ultrasound propagation within the sample \cite{Verbiest}, the cantilever dynamics \cite{Verbiest2,Verbiest3,Bosse,Forchheimer,Turner}, nonlinear mixing \cite{Verbiest4,Platz1,Platz2}, the explicit excitation scheme, the resonance frequency spectrum of the cantilever \cite{Verbiest5,Rabe3}, resonance frequency shifting \cite{Verbiest5}, and the response to variations in the tip-sample interaction \cite{Verbiest5,Parlak,Sarioglu} that are determined by the local elasticity and adhesion of the sample. All these factors can significantly change the heterodyne signal leading to a measurable contrast. Published HFM experiments that provide (at least some) quantitative information are scarce \cite{Cant1} and the actual depth of the subsurface features is confirmed independently only in Ref. \cite{Yam2}.

In this paper, we present a full quantitative analysis that addresses all thinkable, physical contrast mechanisms to explain our experimental observations on a well characterized sample. We show that Rayleigh scattering \cite{Verbiest} would produce a contrast that is orders of magnitude smaller than in the experiment. By calculating the cantilever dynamics for different tip-sample interactions, we show that variations in sample elasticity indeed can lead to contrasts that are, in magnitude, comparable to the experiments. However, we can also rule out this mechanism, as the contrast is inverted with respect to the experimentally observed one. The only remaining possibility is dissipation! As we can also exclude tip damping and ultrasound attenuation, we finally conclude that {\it friction} at {\it shaking} nanoparticles is the responsible physical contrast mechanism. Additional evidence for this comes from an estimate of the involved energy dissipation.

Our analysis shows that the contrast strongly depends on the applied contact force and the precise ultrasonic excitation scheme with respect to the resonance frequencies (and their shifts) of the cantilever.

\section{Results}

\begin{figure}[!b]
\begin{center}
\includegraphics[width=40mm]{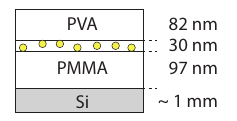}
\end{center}
\caption{{\bf Schematic cross section of the final sample stack:} On the Silicon wafer, we have (from bottom to top), a 97 nm PMMA layer, a 30 nm PVA layer that also contains gold nanoparticles with a diameter of 20 nm, and a 82 nm PVA layer (see Methods and Supplementary Notes 1 and 2 for more details).\label{sfig-1}}
\end{figure}

To enable a quantitative analysis of our measurements, we carefully prepared a sample with 20 nm large gold nanoparticles embedded 82 nm below the surface, see Fig. \ref{sfig-1}. The preparation as well as the independently determined characterization of the sample with AFM, RBS, and SEM is described in Supplementary Notes 1 and 2.

\begin{figure}[!t]
  \includegraphics[width=85mm]{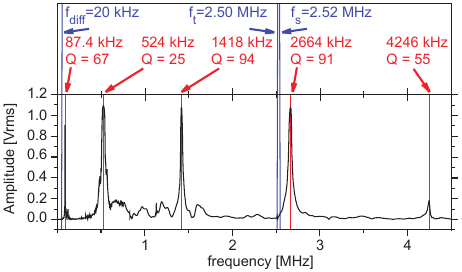}
  \caption{{\bf Experimental excitation scheme:} This scheme falls into the class of {\it off-off resonance} excitation, see Methods. The vibration spectrum of the free hanging cantilever is also shown. A red line indicates a resonance frequency: its value and corresponding Q-factor are indicated in the top panel. The blue lines indicate the applied excitation frequencies of the tip $f_t = 2.50$ MHz, the sample $f_s = 2.52$ MHz, and the difference frequency $f_{diff} = 20$ kHz, which all do not coincide with a resonance frequency of the cantilever.\label{fig1}}
\end{figure}

As the explicit excitation scheme is of crucial importance for the measured HFM contrast, Fig. \ref{fig1} shows our particular experimental choice, called {\it experimental scheme},  with an {\it off-off resonance} excitation scheme (see Methods for the definition of their excitation schemes).

\begin{figure}[!t]
  \includegraphics[width=85mm]{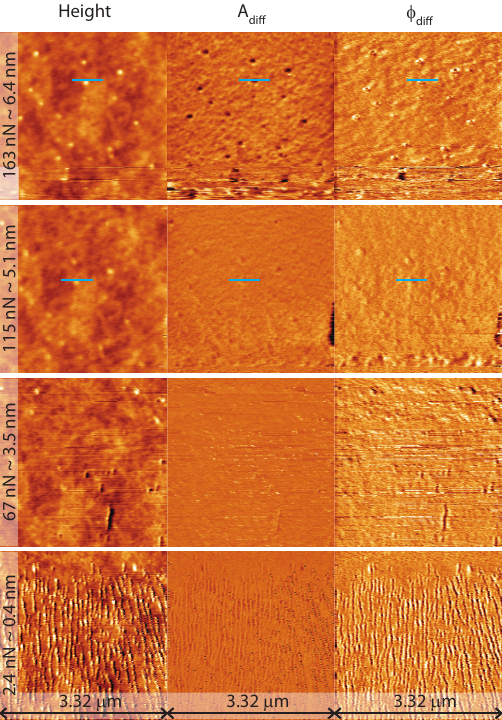}
  \caption{{\bf HFM measurements for different contact forces:} From left to right measured simultaneously: the height and both the amplitude $A_{diff}$ and the phase $\phi_{diff}$ of the difference frequency. The contact force $F_c$ as well as the resulting {\it average indentation} into the sample are indicated at the left in the height images. The gold nanoparticles are only visible at a contact force of 163 nN and 115 nN. At these forces, they are not only visible in the subsurface channels, but also in the height image. We `loose' the nanoparticles in all three channels with decreasing force. At a $F_c = 2.4$ nN, we observe that we damaged the surface, while measuring at higher forces. All images within one channel do have the same (color) range such that the contrast for different contact forces can be compared directly. We provide typical cross sections with absolute values of the three channels at the positions of the nanoparticles in Fig. \ref{fig4}.\label{fig2}}
\end{figure}

Figure \ref{fig2} shows the actual HFM experiment with simultaneously measured height, amplitude $A_{diff}$ and phase $\phi_{diff}$ of the difference frequency $f_{diff}$ for various contact forces $F_c$. Feedback was performed in contact mode operation.
The contact force $F_c$ is decreased from top to bottom: 163 nN, 115 nN, 67 nN, and 2.4 nN. The gold nanoparticles are visible in all channels at $F_c = 163$ nN. The observed density of 1.2 particles/$\mu m^2$ fits the independently determined density. Most of the gold nanoparticles are still visible at $F_c = 115$ nN, although the contrasts are significantly reduced. At lower forces, we do not (or just barely) detect any nanoparticles in any of the channels, which supports the RBS measurements that the gold nanoparticles are indeed fully buried under a 82 nm thick PVA layer. Considering the tip indentation depths (note that this is different from the total height variation, see information at the left side in Fig. \ref{fig2}) and the thickness of the PVA top layer, we have to conclude that the gold nanoparticles are only visible by poking hard enough into the sample, although it is striking that we see them at all in the height images. At $F_c = 2.4$ nN, we start probing the attractive part of the tip-sample interaction and recognize that we have damaged the surface, while measuring earlier at higher contact forces.

At $F_c = 2.4$ nN, both subsurface channels (amplitude and phase) show a clear correlation with the height. As the cantilever mainly probes the attractive part of the tip-sample interaction during an oscillation, the effective contact area of the tip depends on the height variations of the sample: it is much smaller on a mountain than in a valley. Adhesion is directly proportional to the contact area and a variation of it indeed leads to a variation in both the amplitude and the phase of the subsurface signal \cite{Verbiest4}. We conclude that variations in the adhesion do generate a contrast in the subsurface channels.

\begin{figure}[!t]
\begin{center}
\includegraphics[width=85mm]{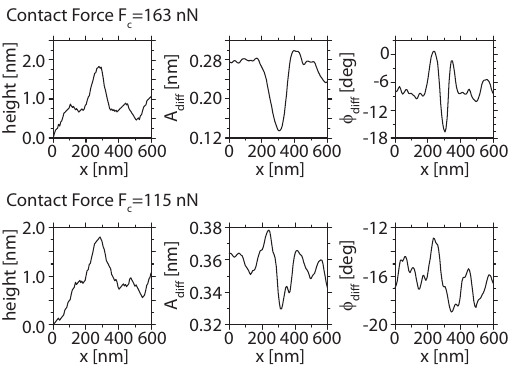}
\end{center}
\caption{{\bf Cross sectional lines of the height, $A_{diff}$ and $\phi_{diff}$ at the position of the blue lines in Fig. \ref{fig2}:} The top panels are for a contact force of 163 nN, whereas the bottom ones are for 115 nN. For a given contact force, the blue lines in Fig. \ref{fig2} are exact on the same location. As all channels are recorded simultaneously, the same pixel in the different channels is taken at exactly the same time. At a contact force of 163 nN, all three channels clearly show strong contrasts, whereas at 115 nN the contrasts in the subsurface channels B an C are almost of the same size as the corresponding background variations. \label{fig4}}
\end{figure}

To quantify the contrasts of the gold nanoparticles in Fig. \ref{fig2}, we extract from cross sectional lines, as shown in Fig. \ref{fig4}, the average values above the nanoparticles for the different channels with respect to their background, see Tab. 1.

Let us first compare the experimental values with the expected contrast based on Rayleigh scattering \cite{Verbiest}, for which we have to normalize the amplitudes $A_{diff}$ with respect to their background amplitudes $A_b$. At $F_c = 163$ nN, we measure a normalized amplitude contrast, $A_c$, of -~0.44 and a phase $\phi_{diff}$ of 7.2 degrees. At $F_c = 115$ nN, the normalized amplitude contrast is -~0.11 and the phase contrast is 2.9 degrees. Based on Rayleigh scattering, the expected normalized amplitude contrast is $10^{-6}$ and the phase contrast is 0.1 millidegree for a gold particle with a diameter of 20 nm buried 50 nm deep under a polymer (PMMA) \cite{Verbiest}. As the experimentally observed normalized amplitude contrast is 5 orders of magnitude larger (and the phase contrast 4 orders of magnitude) than the theoretically predicted ones, we have to conclude that Rayleigh scattering does {\it not} form a major contribution to the physical contrast mechanism (at least not at MHz frequencies).

Recently, it was elucidated how the heterodyne signal is generated: its magnitude strongly depends on both the applied contact force and the specific characteristics of the tip-sample interaction \cite{Verbiest2,Verbiest3,Verbiest4}. In Supplementary Notes 3 and 4 we show, both experimentally and analytically, that the heterodyne signal depends on the elastic properties of the sample, which is characterized by its Young's modulus $E$. For sufficiently soft samples, the amplitude $A_{diff}$ is proportional to $E$. Let us, in the following, consider elasticity variations in the sample, due to the presence of the nanoparticles, as a possible contrast mechanism.

\begin{figure}[!b]
  \includegraphics[width=85mm]{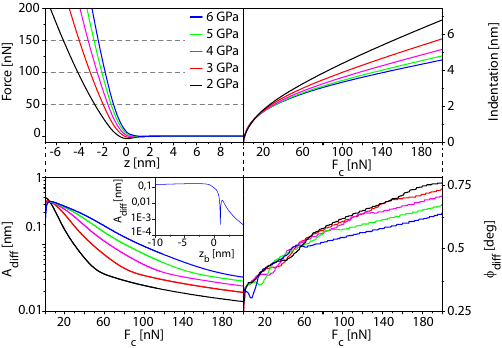}
  \caption{{\bf Results for the \textbf{\textit{experimental excitation}} scheme:} we calculated the tip-sample interaction and, as a function of the applied contact force, the corresponding sample indentation as well as the amplitude $A_{diff}$ and phase $\phi_{diff}$ of the heterodyne signal for different sample elasticities: 2 GPa (black), 3 GPa (red), 4 GPa (magenta), 5 GPa (green), and 6 GPa (blue). The inset in the lower left panel shows $A_{diff}$ for 6 GPa plotted as a function of the height of the cantilever's base, $z_b$, such that a comparison becomes possible with other calculations \cite{Verbiest2,Verbiest3,Verbiest4}.\label{fig3}}
\end{figure}

From an analytical 1D model, we estimate that the Young's modulus above a gold nanoparticle is $\sim 10\%$ higher than the Young's modulus of PVA, which is 2.4 GPa, see Supplementary Note 5.
To determine the contrast formation based on these elasticity variations, we numerically calculated the motion of the cantilever for different tip-sample interactions using the method outlined in \cite{Verbiest2}. The result is shown in Fig. \ref{fig3}, in which we, for reasons of clarity, only show the approach curves. To receive an upper bound on the contrast and to elucidate the contrast formation effect on the basis of small elasticity variations, we consider Young's moduli between 2 and 6 GPa.
As the specific vibration spectrum of the cantilever has great influence on the results, we first matched the spectrum used in the calculations to that of our experiment, see Supplementary Note 6. We call the particular {\it off-off resonance} excitation scheme that we used in this experiment (see Fig. \ref{fig1}), {\it experimental excitation}. The graphical result, see Fig. \ref{fig3}, show the corresponding tip-sample interactions and, as a function of the applied contact force, the indentations as well as the amplitudes $A_{diff}$ and phases $\phi_{diff}$ of the heterodyne signal at the difference frequency. The contrasts at a certain contact force can now be evaluated from the difference in the signals stemming from different elasticities (colors in the graphs). The indentation contrast decreases with decreasing contact force. The amplitude contrast stays almost constant over a large range (and even increases slightly), before it collapses to zero at very small contact forces. The phase contrast strongly depends on the specific excitation scheme, but always collapses to zero at very small contact forces. The extracted height, amplitude and phase values are listed in Tab. \ref{Tab1}. In addition,
to elucidate the effect of different ultrasonic excitation schemes, we also considered an {\it off-off resonance} excitation, in which both ultrasound signals are midway between two resonance frequencies, as well as an {\it off-on resonance} excitation, see Supplementary Note 7. These results are, in addition, tabulated in Tab. \ref{Tab1} for comparison.

\begin{table}[!t]
\begin{tabular}{|c|c|c|c|c|c|c|}
\hline
{\bf method} & $F_{contact}$ & height & $\Delta A_{diff}$ & $A_{b}$ & $A_{c} =$ & $\Delta\phi_{diff}$\\
 & [nN] & [nm] & [pm] & [pm] & $\Delta A_{diff}/A_b$ & [deg]\\
\hline
\hline
{\bf experiment} & {\bf 163} & {\bf 2.8} & {\bf -120} & {\bf 270} & {\bf -0.44} & {\bf 7.2}\\
 & {\bf 115} & {\bf 1.2} & {\bf -40} & {\bf 360} & {\bf -0.11} & {\bf 2.9}\\
\hline
\hline
{\bf exp. scheme} & 163 & 0.08 & 0.87 & 17 & 0.05 & 0.027\\
($2.4 \rightarrow 2.6$ GPa) & 115 & 0.03 & 1.1 & 21 & 0.05 & 0.008\\
\hline
\hline
{\bf exp. scheme} & 163 & 1.8 & 17 & 15 & 1.1 & 0.120\\
($2 \rightarrow 6$ GPa) & 115 & 1.2 & 32 & 19 & 1.7 & 0.083\\
\hline
\hline
{\bf off-off resonance}$^\star$ & 163 & 1.8 & 42 & 20 & 2.1 & 0.014\\
($2 \rightarrow 6$ GPa) & 115 & 1.3 & 63 & 24 & 2.6 & -0.002\\
\hline
\hline
{\bf off-on resonance} & 163 & 1.8 & -0.86 & 7.0 & -0.12 & 11\\
($2 \rightarrow 6$ GPa) & 115 & 1.2 & -0.23 & 9.6 & -0.02 & 12\\
\hline
\multicolumn{7}{l}{$^\star$the ultrasound signals are midway between two resonance frequencies}
\end{tabular}
\caption{{\bf Comparison between experimentally determined and analytically predicted values:} The obtained contrasts in the height, the amplitude $A_{diff}$, the normalized amplitude $A_c$ (for which we also provide the background amplitude $A_b$), and the phase $\phi_{diff}$ for a contact force of 163 nN and 115 nN. The contrasts are obtained from different numerical calculations taking into account specific excitation schemes, see Supplementary Note 7. To receive clear upper estimates, we determined (most of) the contrasts from the differences in the curves of Fig. \ref{fig3} between a sample with 2 GPa and 6 GPa. For completeness, we provide, for the {\it experimental} scheme, also the contrasts obtained from the difference in samples with 2.4 GPa (PVA) and 2.6 GPa (effective elasticity above the nanoparticles, as derived in Supplementary Note 5).\label{Tab1}}
\end{table}

The {\it experimental} scheme with 2.4 GPa (PVA) to 2.6 GPa (effective elasticity above the nanoparticles, see Supplementary Note 5) perfectly reflects both the sample and the measurement conditions. To receive clear upper bounds, we determined further all excitations schemes from the differences between a sample with 2 GPa and 6 GPa.
Starting with the height contrast, we find comparable values between the experiment and the calculated excitation schemes, except for the {\it experimental scheme}  $2.4 \rightarrow 2.6$ GPa.
The decrease in height contrast for smaller contact forces $F_c$ is reproduced for all cases.
Considering the amplitude contrast $\Delta A_{diff}$, the absolute values in the experiment are up to 100 times larger than the calculated ones. One notices three striking issues when performing a more detailed comparison. Firstly, the values of the {\it off-on resonance} case are significantly lower than most other values. This is due to its particular excitation scheme (see Supplementary Note 7), in which the ultrasonic tip amplitude significantly decreases when the cantilever gets into contact with the surface, due to the related frequency shift of the $4^{th}$ resonance. The size of this shift and, therefore, also of the amplitude reduction of the ultrasonic tip vibration, increases both with increasing contact force and with sample stiffness. This excitation scheme with its particular behavior is special, as the frequency shift acts as an amplifier/attenuator to the measured signal. Secondly, in contrast to the experiment, both the {\it off-off resonance} case and the {\it experimental} schemes show a larger contrast at lower force. Although this already indicates a problem, the most striking issue is the sign of the contrast, which is inverted in comparison with the experiment! In any case, the (visco)elasticity above the nanoparticle is for sure {\it increased}, which theoretically leads to a {\it higher} amplitude $A_{diff}$ (see Fig. \ref{fig3} and Supplementary Note 4) and, therefore, to a {\it positive} amplitude contrast $\Delta A_{diff}$. We have to conclude that, although elasticity variations produce a contrast with similar magnitude than in the experiments, they cannot explain the inverted contrast. Consequently, a different physical mechanism must be present.

Please note that the amplitude contrast inversion of $\Delta A_{diff}$ in the {\it off-on resonance} case is due to its particular excitation scheme with the frequency shift of the $4^{th}$ mode. Above the nanoparticle, the amplitude reduction of the ultrasonic tip vibration $A_t$ is significantly larger than the reduction on the PVA without nanoparticles (see Supplementary Note 7). This indicates the importance of the precise excitation scheme and the spectrum of the cantilever for each published HFM measurement, in order to understand it quantitatively.

%
For the sake of completeness, we shortly turn our attention also to the phase behavior. The magnitude of the experimentally observed phase contrast $\Delta \phi_{diff}$ is only comparable to the special case of the {\it off-on resonance} excitation scheme. The large phase shift in this scheme is due to the frequency shift of the $4^{th}$ resonance: the particular {\it off-on resonance} excitation scheme makes the tip vibration especially sensitive to phase changes based on frequency shifts.
Although much smaller in magnitude, a similar argument holds also for the phase shifts in the {\it off-off resonance} and {\it experimental excitation} schemes. Since the ultrasonic tip excitation in the {\it experimental} scheme is closer to the $4^{th}$ resonance frequency of the cantilever, we observe a larger phase contrast than in the {\it off-off resonance} scheme where the excitation of the tip is midway between resonance frequencies.

Summarizing this part, we conclude that the contrast from (small) variations in the sample elasticity results in a much larger contrast than Rayleigh scattering: the order of magnitude is comparable to the experiments.
However, variations in sample elasticity {\it cannot} be the physical contrast mechanism in our HFM experiment, as it would imply an {\it opposite} sign.

\section{Discussion}

Ruling out both variations in the tip-sample interaction (elasticity and adhesion) and Rayleigh scattering, the remaining physical contrast mechanism must lead to a significant reduction of the tip amplitude $A_t$ or the sample amplitude $A_s$ above the nanoparticles, as $A_{diff} \sim A_t A_s / \sqrt{A_t^2+A_s^2}$ \cite{Verbiest4}. These reductions can be described as tip or sample damping.
Tip damping can also be excluded, as it has been surprisingly shown that $A_t$ keeps $99.7\%$ of its amplitude at a contact force of 25 nN even on a hard sample like Si \cite{Verbiest3}. Please note that the damping of the resonance frequencies of a cantilever that is in contact with a sample, is generally assumed to be directly proportional to the Young's modulus of the sample \cite{Santos}.
Without significant tip damping, the contrast must be due to a reduction in the sample amplitude.
Since a reduction of $A_s$ is expected to occur also on the polymer without nanoparticles, and since $A_{diff}$ is larger above the nanoparticle due to the increase in the effective Young's modulus, we need a mechanism that leads to a strong decrease of $A_s$ {\it only} above the nanoparticle to overcompensate the increase in $A_{diff}$ such that it effectively leads to a contrast inversion (holes in $A_{diff}$, see Fig. \ref{fig2}).


Let us start with a possible vertical motion of the nanoparticles in the polymer matrix. At low ultrasonic sample frequencies, this motion is surely in phase with the excitation. However, if the ultrasonic excitation is above the resonance frequency of the system ``nanoparticle in polymer", the motion will be out of phase leading to a significant reduction of $A_s$ only above the nanoparticles. The problem is, however, that the sample excitation is at 2.5 MHz and that we estimate the resonance frequency of the ``nanoparticle in polymer" system to be $\sim 2.2$ GHz (see Supplementary Note 8). The nanoparticles should, therefore, simply follow the ultrasonic displacements of the polymer.

Another mechanism worth considering is sample damping (reduction of $A_s$) by energy dissipation at the nanoparticles. Next to contrast formation based on attenuation or friction, a temperature effect might additionally enhance the contrast, especially if the elasticity of the polymer would have a strong temperature dependence. Therefore, we estimate the energy dissipation from the experiment. We determine the sample amplitude $A_s$ (far away from the nanoparticle) in analogy to the method described in \cite{Verbiest2} (see also Methods). With $A_s \sim 0.22$ nm at $F_c = 163$ nN, we need a reduction of $\sim 41\%$ to explain the observed contrast. $A_s$ is $\sim 0.29$ nm at $F_c = 115$ nm, which corresponds to a reduction of $\sim 13\%$.
Both estimations deliver a similar value: 0.83 and 0.37 pW, respectively. This breaks down to an energy dissipation at the nanoparticles of less than 2.07 eV/oscillation of the ultrasonic sample excitation.
This value is so small that we can rule out also any temperature effects. The only remaining physical mechanism that might cause this energy dissipation is ultrasound attenuation {\it within} the nanoparticles as well as friction at the interface {\it between} the nanoparticles and the polymer.

The ultrasound attenuation for gold is $\sim 150$ times {\it smaller} than the attenuation for PVA. Therefore the total energy dissipation is {\it less} at the positions measured above the nanoparticles than at the positions far away from them. This effect results, in comparison to the experiment, again in a wrong sign of the contrast, as $A_s$ should be larger above the nanoparticles. We estimate this resulting energy `gain' based on a smaller ultrasound attenuation at the nanoparticles to be 0.45 eV/oscillation. The dissipation that causes the contrast, must be increased with this value to overcompensate it and lead to contrast inversion.

This means that we are left with friction at the interface between the nanoparticles and the PVA. Due to a weak (chemical) bonding between the gold and the PVA, the nanoparticles might (slightly) slip instead of following all displacements of the PVA. One might even consider a small cavity around the nanoparticles such that they are {\it shaken} up and down. Both effects would lead to a significant amount of friction at the interface. Considering {\it shaking} nanoparticles, we are able to explain our observed contrast with a total energy dissipation of 2.52 eV/oscillation at the nanoparticles. This value is comparable (and definitively in the right order of magnitude) with the energy dissipation derived from atomic scale friction experiments of a sharp tip in contact with a surface \cite{Dienwiebel}. Note that the tip radius in these experiments is comparable to the radius of the nanoparticles.

Pinpointing the physical mechanism to friction at {\it shaking} nanoparticles, we can consider the consequences for the lateral resolution. If one assumes that the propagation in amplitude reduction obeys a scattering-like behavior, the `fingerprints' of the nanoparticles at the surface should show a significantly larger diameter than the diameter of the nanoparticles. Moreover, as we are measuring in near-field, the size of the `fingerprints' should be in the order of the depth of the nanoparticles. The deeper the nanoparticle is, the larger should be its image at the surface. These considerations stand in clear contrast to experimental observations: nanoparticles with a diameter of $\sim 17.5$ nm, buried 500 nm deep, are imaged with a diameter of only $20$ nm \cite{Dra1}, and the imaged fingerprint is even decreasing with increasing depth of the nanoparticles \cite{Yam2}. A solution to this might be found by considering a combination of a stress field that is introduced on the nanoparticle by the indenting tip \cite{Parlak}, a resulting {\it shaking} that is no longer parallel to the initial ultrasonic displacements of the PVA, and a highly anisotropic propagation of the amplitude reduction such that there is a significant enhancement in the direction of the {\it shaking} movement.

\section{Acknowledgements}

We gratefully thank Prof. R. W\"{o}rdenweber and E. Hollmann (Forschungszentrum J\"{u}lich, Germany) for the RBS measurements and a first analysis of the data, as well as M.Y. Yorulmaz for assistance with the sample preparation. The research described in this paper has been performed under and financed by the NIMIC \cite{NIMIC} consortium under project 4.4. T.H. Oosterkamp acknowledges support from an ERC starting grant.

\section{Author contributions}

The project was initiated and conceptualized by M.J.R. G.J.V. performed all the measurements, the simulations, and the analytical calculations presented in this study. G.J.V. and M.J.R. interpreted the results and narrowed down the number of possible physical contrast mechanisms. T.H.O. helped with the interpretation and suggested ``Friction at shaking nanoparticles'' as a possible contrast formation mechanism. G.J.V. and M.J.R. wrote the manuscript together, which was carefully read and improved by all authors.

\section{Methods}

As a quantitative analysis of the contrast mechanism is impossible without a well-characterized sample, we carefully prepared a stack consisting of the following layers (from bottom to top): a Si wafer with native oxide, a $\sim 97$ nm thick PMMA layer, a 30 nm thick PVA layer with embedded gold nanoparticles (diameter 20 nm), and a 82 nm thick PVA top layer. The density of the gold nanoparticles was determined via AFM and SEM to be $0.7 \pm 0.6$ particles/$\mu m^2$. The precise sample preparation as well as its detailed characterization, in which we even determined the depth of the Au nanoparticles with an independent measurement based on Rutherford backscattering, is described in detail in Supplementary Note 1 and 2.

In our HFM experiment, we chose the ultrasonic excitation frequencies of both the tip and the sample as well as the difference frequency {\it off} resonance, i.e. not on (or within the width) of a resonance peak of the cantilever.
We call this excitation scheme {\it off-off resonance}. The first {\it on/off} indication describes whether $f_{diff}$ (heterodyne signal) is tuned to a resonance frequency of the cantilever, whereas the second {\it on/off} indication describes whether $f_t$ (ultrasonic tip excitation) is tuned to a resonance. This leads to four different excitation schemes, of which we evaluate also the {\it off-on} scheme in more detail in Supplementary Note 7.

Figure \ref{fig1} shows the excitation scheme and the vibration spectrum of the free hanging cantilever, of which we calibrated the spring constant to be 2.7 N/m using the thermal noise method \cite{Hutter}.
Using the method described in \cite{Verbiest3,Verbiest4}, we determined the ultrasonic tip amplitude to be $A_t = 1.34$ nm and the ultrasonic sample amplitude to be $A_s = 0.37$ nm.

\newpage
\onecolumngrid

\section*{Supplementary Notes}

\subsection{1. Sample Preparation\label{SSec1}}

\begin{figure}[!ht]
\begin{center}
\includegraphics[width=120mm]{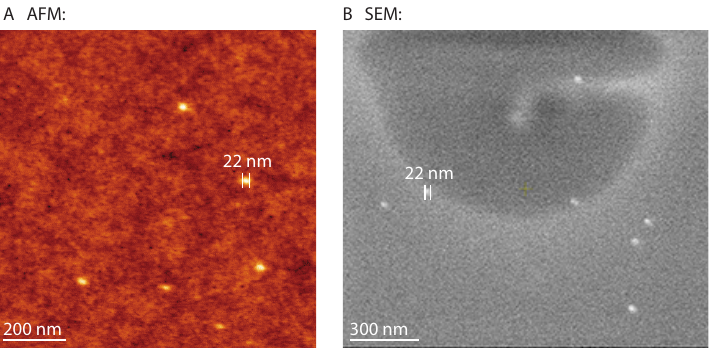}
\end{center}
\caption{{\bf Density and distribution of the nanoparticles.} (A) measured with an AFM before burying them under an additional PVA layer, and (B) measured with a SEM on the final sample, where the nanoparticles are covered by a 82 nm thick top PVA layer.\label{sfig0}}
\end{figure}

Inspired by the sample with buried gold nanoparticles of Shekhawat and Dravid \cite{Dra1b}, we set out to produce comparable ones. We decided to use gold nanoparticles with a diameter of 20 nm ($\pm 10\%$), which we got from BBI Solutions \cite{BBIb}. A schematic cross section of the final sample that we used in the current study, is shown in Fig. 1 in the main text. In the following, we describe important issues of the sample preparation and provide the recipe.
As a substrate, we used a freshly with acetone cleaned Silicon (100) wafer that was covered with a native oxide. The polymer layers (including the suspension with the nanoparticles) were deposited by means of a spin coater, see also the recipe below. We decided to use two different polymers: polymethylmethacrylaat (PMMA) with a degree of polymerization of 970 and polyvinyl alcohol (PVA) with a degree of polymerization of 2700. The degree of polymerization is the number of monomers in the molecule and it characterizes the length of a single polymer molecule. This information is important, as the material properties of the polymer layers strongly depend on the molecule length. As we faced some problems with clustering of the nanoparticles as well as with their density, we describe these issues shortly in the following. Our first attempt to create a layer of gold nanoparticles on top of a spin coated PMMA layer was to let a suspension of pure (Milli-Q) water with gold nanoparticles evaporate at ambient conditions. This led to large ``mountains'' of clustered nanoparticles, which we measured with an Atomic Force Microscope (AFM). In our second attempt, we tried to {\it embed} the gold nanoparticles within a PVA layer by dissolving them in a PVA solution before spin coating on top of the PMMA. The gold nanoparticles did stick out with their ``heads'' just above the PVA layer, with which they were simultaneously spin coated, such that we easily could verify the density, again, with AFM. This approach resulted in an (for our research) unsuitable low density of nanoparticles of less than 0.1 nanoparticle/$\mu$m$^2$. By increasing the concentration of the nanoparticles in the PVA solution, we were able to increase the density to $0.7 \pm 0.6$ nanoparticle/$\mu$m$^2$. We derived this distribution from AFM measurements, see Fig. \ref{sfig0}A. Finally, we buried the nanoparticles by spinning another PVA layer on top of this structure. As we considered that the solvent, which is present while spinning the additional PVA layer, might (partially) dissolve the thin nanoparticles/PVA layer that is to be buried, leading to a possible redistribution of the nanoparticles, we counterchecked the density with a Secondary Electron Microscope (SEM) on the final sample with the top PVA layer. Due to the different electron emissions between gold and PVA, the SEM is capable of imaging the nanoparticles, even if they are buried under a 82 nm thick PVA layer, see Fig. \ref{sfig0}B.

The final recipe for the sample production is as follows:

\begin{enumerate}
\item solution: 30 mg PMMA / mL Toluene\\
This results in a $\sim 97$ nm thick PMMA layer, which was confirmed independently with an AFM measurement \cite{Yorulmazb}.
\item solution: 250 $\mu$L of 2 mg PVA / mL water + 750 $\mu$L suspension of pure water and gold nanoparticles\\
This leads to a PVA layer with embedded gold nanoparticles with a diameter of 20 nm. The thickness of this PVA layer is less than 30 nm ($\sim$ 10 nm), as we verified with AFM that the ``heads'' of the gold nanoparticles are sticking out. After burying this PVA layer with the top PVA layer, we find an effective thickness of 30 nm for this layer that contains the nanoparticles.
\item solution: 2mg PVA / mL water\\
This step leads to a $\sim 82$ nm thick PVA layer.
\end{enumerate}

Each step in the recipe represents an individual spin coating procedure. In each spin coating step, a droplet of the corresponding solution was put onto the sample by means of a pipet before the spin coater started to rotate for

\begin{description}
\item[] 5 s at 2000 rpm immediately followed by
\item[] 90 s at 4000 rpm.
\end{description}

Although we did not apply explicitly a curing (baking) step of the final sample after the preparation, the complete sample was baked for approximately 3 minutes at $\sim 140$ $^0$C to glue it with crystalbond 509 onto the ultrasonic transducer of the sample. This was always done within 24 hours after the spin coating procedure. We assume that, during this baking procedure, most of the remaining solvents in the sample were evaporated.

\subsection{2. Independent Verification of the Nanoparticle Depth\label{SSec8}}

\begin{figure}[!ht]
\begin{center}
\includegraphics[width=85mm]{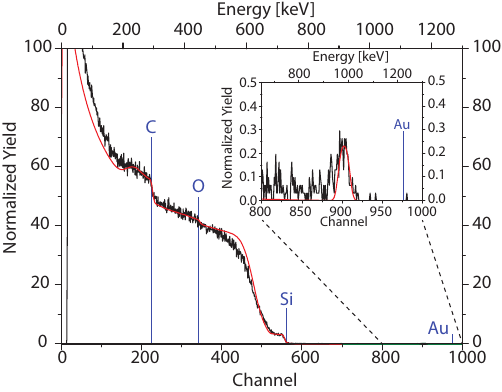}
\end{center}
\caption{{\bf The RBS measurement (black) and the corresponding result of the simulation (red).} The surface channel of the different elements in our sample (C, O, Si, and Au) are indicated in blue. Although almost at the detection limit of the RBS setup, the inset clearly shows a signal obtained from the buried gold nanoparticles.\label{sfig8}}
\end{figure}

In order to quantify HFM experiments, it is of great importance to have a well defined sample, in which the depth of the subsurface particles (or features) is counterchecked with an independent technique. To this end, we performed a Rutherford Backscattering Spectrometry (RBS) measurement on the sample that we used for experiments, to quantify the exact depth of the gold nanoparticles as well as the thickness of the individual layers. To deduce quantitative data from an RBS measurement, it is necessary to perform a simulation \cite{RUMPb}. Figure \ref{sfig8} shows both the RBS measurement (black) and the corresponding result of the simulation (red). The surface channels of the different elements in our sample (Carbon, Oxygen, Silicon, and Gold) are indicated in blue. Although almost at the detection limit of the RBS setup, the inset clearly shows a signal obtained from the buried gold nanoparticles: it is a sharp distribution, which indicates a well defined depth of the nanoparticles, with a clear shift away from the surface channel of Au, from which we can determine the thickness of the top PVA layer.

\begin{table}[!ht]
\begin{tabular}{|c|c|c|c|ccccc|}
\hline
layer & ``material'' & thickness & density & C & H & O & Au & Si\\
 & & [nm] & [$10^{23}$ atoms/cm$^3$] &  &  &  &  & \\
\hline
1 & PVA & 17 & 1.254 & 5.000\,&\,4.000\,&\,0.500\,&\,0.000\,&\,0.000 \\
2 & PVA & 65 & 1.254 & 5.000\,&\,4.000\,&\,0.500\,&\,0.000\,&\,0.430 \\
3 & PVA & 30 & 1.254 & 5.000\,&\,4.000\,&\,0.500\,&\,0.001\,&\,0.430 \\
4 & PMMA & 97 & 1.083 & 12.000\,&\,8.000\,&\,1.000\,&\,0.000\,&\,0.800 \\
5 & Si & 3000 & 4.979 & 0.000\,&\,0.000\,&\,0.000\,&\,0.000\,&\,1.000 \\
\hline
\end{tabular}
\caption{{\bf Layer thickness and composition according to our simulation that fit the RBS measurements best.} Each layer is specified by its thickness [nm], its density [$10^{23}$ atoms/cm$^3$], and its composition C:H:O:Au:Si (not normalized to 1, as this is performed automatically by the used software). \label{sRUMP}}
\end{table}

The RBS spectrum in Fig. \ref{sfig8} shows that Si is present just below the sample's surface, see the rise (and the tiny plateau) in the spectrum almost at the Si surface channel as well as layer 2 in Tab. \ref{sRUMP}. We can explain this with the presence of air bubbles in our sample and/or holes in some of the spin coated polymer layers. As a consequence, the best simulation result contains 4 layers on top of the Si wafer (see Tab. \ref{sRUMP}).

The combined thickness of layers 1 and 2 is 82 nm. Therefore, the gold nanoparticles are buried approximatly 82 nm below the surface.
The thickness of the underlying PMMA layer is approximately 97 nm. We verified the total thickness of 209 nm by scanning over scratches on the sample with an AFM. The minimum thickness that we found in all AFM heightlines is $\sim 250$ nm, which confirms the RBS analysis.

It is striking that the sample contains more C than expected, but less O. From the simulation, we find the following composition for the PVA: $C_2 H_{1.6} O_{0.2}$, which has to be compared to $C_2 H_{4} O_{1}$. The lack of oxygen can be explained either by the formation of water during the baking procedure at $\sim 140$ degrees $^0$C after the spin-coating or by a decomposition of the polymer layers during the RBS measurements (a clear spot on the sample surface was visible after the experiment). For the PMMA layer we find a composition of $C_5 H_{3.3} O_{0.4}$ instead of $C_5 H_{8} O_{2}$.

\subsection{3. Experimental Dependence of the Difference Frequency Amplitude $A_{diff}$ on the Sample Elasticity\label{SSec2}}

To experimentally address the dependence of the amplitude $A_{diff}$ of the heterodyne signal at the difference frequency on the elasticity of the sample, which is characterized by its Young's modulus $E$, we present results for the difference frequency generation on both a soft sample ($\sim 97$ nm thick PMMA, $E \sim 2.4$ GPa) and a hard sample (Si(100) wafer, $E \sim 179$ GPa). The HFM experiment was performed with a similar cantilever as described in the main text.

We obtained the Young's modulus on PMMA by fitting an experimentally obtained tip-sample interaction $F_{ts}$ with the Derjaguin-Muller-Toporov (DMT-) model \cite{Derjaguinb}. A parameter called $\lambda$, which is related to the elasticities of the tip and the sample, is usually used to differentiate between the applicabilities of different models that describe the tip-sample interaction \cite{Johnsonb}. As $\lambda = 0.63$ in our case, one should use the Maugis-Dugdale model \cite{Maugisb}. Nevertheless, our approach with the DMT-model is fully justified, as we have demonstrated in \cite{Verbiest3b} that it does not matter at all for the numerical simulations which of the models describes the tip-sample interaction, as long as the fit perfectly matches the (experimentally obtained) tip-sample interaction. The only thing that matters is the particular shape (form) of $F_{ts}(z)$ and {\it not} the model that is used to describe this particular interaction.

The cantilever has a spring constant of $2.0 \pm 0.4$ N/m, which was calibrated using the thermal noise method \cite{Hutterb}. We applied an {\it off-off} resonance excitation scheme with an ultrasonic tip frequency of 2.870 MHz and an ultrasonic sample frequency of 2.871 MHz leading to a heterodyne signal at a difference frequency $f_{diff}$ of 1 kHz. The ultrasonic vibration amplitudes of both the tip $A_t$ and the sample $A_s$ were slightly different for the two experiments: $A_t = 0.94$ nm and $A_s = 0.32$ nm on Si, whereas $A_t = 1.23$ nm and $A_s = 0.18$ nm on PMMA. The tip amplitudes were determined using the procedure outlined in \cite{Verbiest3b} and below we describe how we determined the sample amplitudes from the measurements. We measured the amplitude $A_{diff}$ of the difference frequency as a function of the cantilever's base position $z_b$ on both the Si and the PMMA layer. $z_b$ is defined such that $z_b = 0$, if the deflection $\delta = 0$ during the approach cycle of the cantilever to the surface. This is exactly the point, at which the {\it effective} interaction on the tip changes sign from an attractive interaction to a repulsive interaction.

\begin{figure}[!ht]
\begin{center}
\includegraphics[width=120mm]{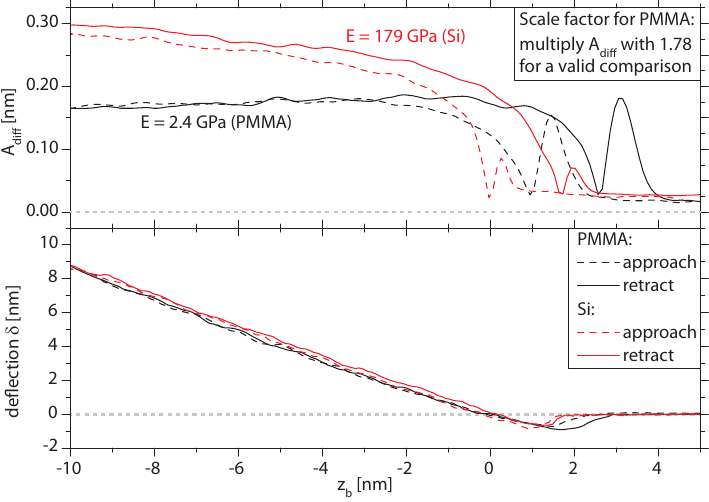}
\end{center}
\caption{{\bf Experimental Dependence of $A_{diff}$ on the Sample Elasticity.} Top panel: the amplitude $A_{diff}$ of the heterodyne signal as a function of the cantilever's base position $z_b$ on both a hard Si sample (red) and a soft PMMA layer (black). As $A_{diff} \propto A_s \cdot A_t/\sqrt{A_s^2+A_t^2}$, see below and \cite{Verbiest4b}, and as the vibration amplitudes are slightly different for the measurements on Si and PMMA, one has to multiply the amplitude $A_{diff}$ for the PMMA case with a scale factor of 1.78 to accommodate for a valid comparison. Please note that, even with this correction factor, $A_{diff}$ is significantly larger on the hard Si surface (179 GPa) than on the soft $\sim 97$ nm thick PMMA layer (2.4 GPa). The lower panel shows the corresponding deflection $\delta$ of the cantilever.\label{sfig1}}
\end{figure}

Figure \ref{sfig1} shows the amplitude $A_{diff}$ of the heterodyne signal as well as the corresponding deflection of the cantilever as a function of the cantilever's base position $z_b$ on both a Si sample (red) and a $\sim 97$ nm thick PMMA layer (black). To determine the ultrasonic vibration amplitude $A_s$ of the sample, we can estimate $A_s$, for the case of Si, from the height of the plateau, using the method described in \cite{Verbiest3b}, to be 0.32 nm. Without the existence of a clear plateau in the PMMA case (note the slight decrease of $A_{diff}$ for negative $z_b$), we instead use the maximum amplitude of the difference frequency $A_{diff}$ (0.18 nm) for the estimation and find $A_s = 0.18$ nm.

To enable a valid comparison between the measurements on the two different samples, one has to multiply the amplitude $A_{diff}$ for the PMMA case with a correction factor of 1.78, as $A_{diff} \propto A_s \cdot A_t/\sqrt{A_s^2+A_t^2}$ (see below and \cite{Verbiest4b}) and as the vibration amplitudes are slightly different for the measurements on the different samples. Taking this correction factor into account, one still observes that, for the same contact force, $A_{diff}$ is significantly larger on the hard Si surface than on the soft $\sim 97$ nm thick PMMA layer. The peaks in the attractive regime are larger for the soft PMMA sample, because the adhesion is larger on the PMMA sample than on the Si sample (please note the difference in deflection in the attractive part of the tip-sample interaction). Thus we conclude that the amplitude $A_{diff}$ significantly depends on the elasticity of the sample and increases with increasing Young's modulus E.

\subsection{4. Analytical Dependence of the Difference Frequency Amplitude $A_{diff}$ on the Sample Elasticity\label{SSec3}}

Recently, an analytical theory has been developed that completely describes the generation of the heterodyne signal at the difference frequency for HFM experiments \cite{Verbiest4b}. The signal is characterized by the following analytical expressions:

\begin{align}
A_{diff}e^{{\rm i}\phi_{diff}} &= \frac{A_s A_t}{\sqrt{A_s^2+A_t^2}} \frac{I_2 e^{{\rm i} (\phi_s-\phi_t)}}{|H^{-1}(\omega_{diff})| e^{{\rm i} \Lambda} - I_1}\label{s2-e1}\\
I_1 &= \frac{1}{\pi}\int_{-1}^1 \frac{\partial F_{ts}}{\partial z}\left(z_b + \delta + \sqrt{A_s^2+A_t^2}u\right) \frac{du}{\sqrt{1-u^2}}\label{s2-e2}\\
I_2 &= \frac{\sqrt{A_s^2+A_t^2}}{2\pi}\int_{-1}^1 \frac{\partial^2 F_{ts}}{\partial z^2}\left(z_b + \delta + \sqrt{A_s^2+A_t^2}u\right)\sqrt{1-u^2}du\label{s2-e3}
\end{align}

, in which $A_{diff}$ and $\phi_{diff}$ are the amplitude and the phase, respectively, of the signal at the difference frequency, and $A_s$ and $A_t$ are the ultrasonic vibration amplitudes of the sample and the tip with corresponding phases $\phi_s$ and $\phi_t$. $|H^{-1}(\omega_{diff})|$ represents the absolute value of the inverse transfer function and its corresponding phase shift $\Lambda$, $F_{ts}$ is the tip-sample interaction as a function of the tip-sample distance z, $z_b$ is the position of the cantilever's base, and $\delta$ is the deflection of the cantilever.

The integrals $I_1$ and $I_2$ completely determine the generation of the signal at the difference frequency and they both depend on the particular tip-sample interaction $F_{ts}$. As the tip-sample interaction in the experiment can be best described by the DMT-model \cite{Derjaguinb,Verbiest3b}, $F_{ts}$ can be expressed by

\begin{equation}                                                                                                                                     F_{ts}(z) =                                                                                                                                          \begin{cases}                                                                                                                                        \displaystyle -\frac{HR}{6a_0^2}+\frac{4}{3} E_f \sqrt{R}(a_0-z)^{3/2} & \text{if } z \leq a_0,\\
&\\
\displaystyle -\frac{HR}{6z^2} & \text{if } z > a_0.                                                                                                             \end{cases}\label{s2-e4}                                                                                                                               \end{equation}

, in which $R$ is the radius of the cantilever's tip, $H$ the Hamaker constant, $a_0$ the distance at which the repulsive part of the tip-sample interaction is first felt by the cantilever ($\sim$ at the minimum of $F_{ts}$), and $E_f$ is an effective Young's modulus describing the effective tip-sample stiffness. This effective Young's modulus $E_f$ is determined by the elasticities ($E_t$ and $E$) as well as the Poisson ratio's ($\mu_t$ and $\mu$) of the cantilever and the sample, respectively:

\begin{equation}
\frac{1}{E_f} = \frac{1 - \nu^2}{E} + \frac{1 - \nu_t^2}{E_t}\label{s2-e5}
\end{equation}

Since we probe our final sample that consists of several polymers layers, of which $E \sim 2.4$ GPa, with a hard Silicon cantilever with $E_t \sim 179$ GPa, we can neglect $(1 - \nu_t^2)/E_t$ and receive that the effective elasticity $E_f$ is directly proportional to the elasticity $E$ of the sample. The repulsive part of the tip-sample interaction, see Eq. \ref{s2-e4}, is, therefore, also directly proportional to the elasticity $E$ of the sample. As a consequence, this is valid also for the integrals $I_1$ and $I_2$ described by Eqs. \ref{s2-e2} and \ref{s2-e3}. Using these proportionality relations in Eq. \ref{s2-e1}, we find a simple expression for the elasticity dependence of the amplitude $A_{diff}$ of the heterodyne signal:

\begin{equation}
A_{diff} \propto \left|\frac{E}{\gamma + E}\right| = \frac{E}{\sqrt{E^2 + |\gamma|^2 + 2 E {\rm Re}\left[\gamma\right]}}\label{s2-e6}
\end{equation}

, in which $\gamma$ is a complex constant. If the cantilever is completely in the Hertzian contact regime ($z < a_0$) during its oscillation, gamma can be written as

\begin{equation}
\gamma = \cfrac{|H^{-1}(\omega_{diff})| e^{{\rm i} \Lambda}}{\left[\frac{2\sqrt{R}\sqrt{A_s^2+A_t^2}}{\pi}\right]{\displaystyle\int_{-1}^1} \sqrt{\frac{\alpha-u}{1-u^2}}du}\label{s2-e7}
\end{equation}

, in which $\alpha$ is the normalized indentation given by:

\begin{equation}
\alpha = \frac{a_0 - z_b - \delta}{\sqrt{A_s^2+A_t^2}}\label{s2-e8}
\end{equation}

We can evaluate an lower estimate for $\gamma$ by setting the normalized indentation $\alpha = 1$ and noticing that for smaller $\alpha$, the integral in the expression for $\gamma$ would become smaller, and $\gamma$, therefore, larger. Using the ultrasonic amplitudes of both the tip and the sample, we appraise $\sqrt{A_s^2+A_t^2} = 1.39$ nm. For the tip radius we assume $R = 5$ nm. The inverse transfer function $|H^{-1}(\omega_{diff})|$ can be derived as described in \cite{Verbiest4b}, in which we take a spring constant of $2.5$ N/m and set $\Lambda$ to zero. This leads to the following estimates for $\gamma$ and $A_{diff}$:

\begin{align}
\gamma &= 0.5\,{\rm GPa}\nonumber\\
A_{diff} &\propto \frac{E\,{\rm [in\,GPa]}}{0.5+E\,{\rm [in\,GPa]}}\label{s2-e9}
\end{align}

, in which $E$ has to be inserted in GPa. Equation \ref{s2-e9} describes an analytical dependence of $A_{diff}$ on the sample elasticity $E$. For soft samples, in which E is smaller than 0.5 GPa, $A_{diff}$ is approximately proportional to E. Therefore, we also expect analytically that a harder sample results in a higher amplitude $A_{diff}$ of the heterodyne signal, especially above the nanoparticles, where the effective elasticity is slightly increased with respect to the soft polymer. On very hard samples, with $E \gg 0.5$ GPa, $A_{diff}$ approaches a constant value and becomes independent of $E$.

Please note that we have neglected the influence of the elasticity on both the deflection of the cantilever and the transfer function of the cantilever. However, this is a valid approximation, as we never saw a decrease in the amplitude $A_{diff}$ of the difference frequency at a given contact force while the elasticity $E$ of the sample was increased.

\subsection{5. Effective Sample Elasticity above the Nanoparticles\label{SSec4}}

In this section, we derive an upper bound for the effective sample elasticity, measured at the sample surface, that is increased by the presence of the buried nanoparticles in the polymer.

\begin{figure}[!ht]
\begin{center}
\includegraphics[width=40mm]{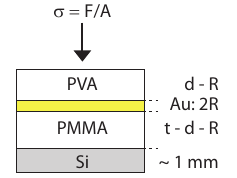}
\end{center}
\caption{{\bf A schematic cross section of the sample.} The sample consists of a PVA layer (PVA), the gold nanoparticle (Au), and a PMMA layer (PMMA). The Au is buried at the depth $d$, and has a radius $R$. The total thickness of the sample is denoted with $t$. The sample is compressed by a stress $\sigma$, which is equal to the force $F$ per unit area $A$.\label{sfig2}}
\end{figure}

Figure \ref{sfig2} shows a schematic cross section of the sample, which consists of a PVA layer (PVA), the gold nanoparticles (Au), and a PMMA layer (PMMA). The Au is buried at the depth $d$, and has a radius $R$. The total thickness of the sample is denoted with $t$. The sample is compressed by a stress $\sigma$, which is equal to the force $F$ per unit area $A$.

From linear elasticity theory, we know that an applied external stress is negatively proportional to the relative change in thickness, in which the proportionality factor is given by the Young's modulus $E$ of the material. For our sample this reduces to the following equations:

\begin{equation}
\sigma = -\frac{E_{PVA}}{d-R}(\delta d - \delta R) = -\frac{E_{Au}}{2R}(2\delta R) = \frac{E_{PMMA}}{t-d-R}(\delta t - \delta d - \delta R)\label{s3-e1}
\end{equation}

, in which $E_{PVA}$, $E_{Au}$, and $E_{PMMA}$ are the Young's moduli of PVA, gold, and PMMA, respectively, $\delta d$ is the variation in depth of the nanoparticle, $\delta R$ is the variation in radius of the nanoparticle, and $\delta t$ is the variation in thickness of the sample.

It is straightforward to derive the solutions for $\delta d$, $\delta R$, and $\delta t$, from Eq. \ref{s3-e1}:

\begin{align}
\delta d & = -R\sigma \left[ E_{Au}^{-1} - E_{PVA}^{-1} \right] - d\sigma E_{PVA}^{-1}\label{s3-e2}\\
\delta R & = -R\sigma E_{Au}^{-1}\label{s3-e3}\\
\delta t & = -t\sigma E_{PMMA}^{-1} -d\sigma \left[ E_{PVA}^{-1} - E_{PMMA}^{-1} \right] -R\sigma \left[ 2 E_{Au}^{-1} - E_{PVA}^{-1} - E_{PMMA}^{-1} \right]\label{s3-e4}
\end{align}

If one introduces an effective Young's modulus $E_{eff}$, the complete sample with all three layers can be regarded also as a sample consisting of one layer with a thickness $t$ of an isotropic material such that

\begin{equation}
\sigma = - \frac{E_{eff}}{t} \delta t\label{s3-e5}
\end{equation}

By substituting Eq. \ref{s3-e4} in Eq. \ref{s3-e5}, we find an expression for the effective Young's modulus $E_{eff}$:

\begin{empheq}[box=\fbox]{equation}
E_{eff} = \frac{1}{E_{PMMA}^{-1} + \tfrac{d}{t} \left[ E_{PVA}^{-1} - E_{PMMA}^{-1} \right] + \tfrac{2R}{t} \left[ E_{Au}^{-1} - 0.5 E_{PVA}^{-1} - 0.5 E_{PMMA}^{-1} \right]}\label{s3-e6}
\end{empheq}

Let us now discuss the two limits of this equation. Firstly, if the diameter $2R$ of the nanoparticle is equal to the thickness $t$ of the sample (and thus $d = 0$), we find that $E_{eff} = E_{Au}$. Secondly, if the radius $R$ of the nanoparticle is equal to zero and the sample is infinitely thick ($t \gg d$), we find that $E_{eff} = E_{PMMA}$. Thirdly, if the radius $R$ of the nanoparticle is equal to zero and $d = t$, we find that $E_{eff} = E_{PVA}$. These results reflect correct expectations, as the sample consists only of Au in the first case, only of PMMA in the second case, and only of PVA in the third case.

Equation \ref{s3-e6} provides an upper bound on the elasticity on the surface above a nanoparticle. In reality, the variation in elasticity due to a nanoparticle should be derived from a 3D calculation, as the stress is spread out also laterally through the sample \cite{Deryuginb}. As a consequence, the rise in elasticity caused by the presence of the nanoparticle decreases with increasing depth of the nanoparticle. This effect is comparable to a stone underneath a pillow: if one just touches the pillow, the stone is not felt, but if one pushes harder into the pillow, the presence of the stone is clearly noticed.

Let us now calculate the expected effective elasticity increase for our samples. PMMA and PVA, both have a similar Young's modulus: $E_{PMMA} \sim E_{PVA} = 2.4$ GPa. Under this assumption, Eq. \ref{s3-e6} reduces to:

\begin{empheq}[box=\fbox]{equation}
E_{eff} = \frac{1}{E_{PVA}^{-1} + \tfrac{2R}{t} \left[ E_{Au}^{-1} - E_{PVA}^{-1} \right]}\label{s3-e7}
\end{empheq}

We assume that the Young's modulus of the gold nanoparticle is equal to that of bulk gold, which is 78 GPa \cite{Ramosb} and consider the total thickness to be $t = 209$ nm. For the radius, we take $R = 10$ nm of the gold nanoparticles, as this is the average of their radii distribution \cite{BBIb,Yorulmazb}. Using this value for the radius $R$, we find the effective Young's modulus $E_{eff}$ to be equal to 2.65 GPa. Therefore, the surface directly above the nanoparticle has (at maximum) a 10\% higher Young's modulus than that of the bulk polymer of 2.4 GPa.

\subsection{6. Setting up the Numerical Calculations\label{SSec5}}

As the amplitude and phase contrast highly depend on both the exact excitation scheme  and precise resonance frequency spectrum of the cantilever, which can even result in a contrast inversion, it is of uttermost importance to match the spectrum of the cantilever in the numerical calculation (numerical cantilever) to the spectrum of the cantilever used in the experiment (experimental cantilever). In this section, we describe the matching procedure.

\begin{figure}[!ht]
\begin{center}
\includegraphics[width=120mm]{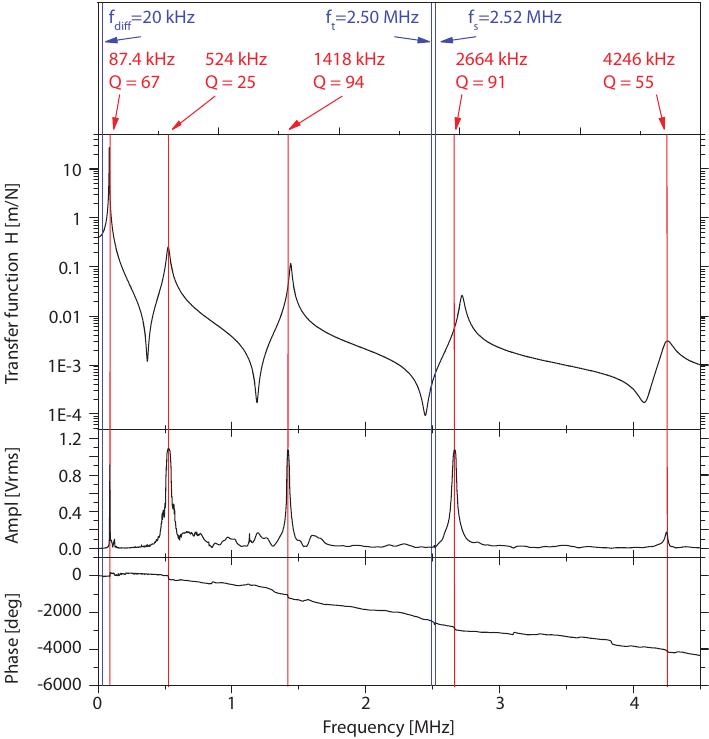}
\end{center}
\caption{{\bf The vibration spectrum of both the {\it experimental} and the {\it numerical} cantilever.} The bottom two panels show the amplitude and the phase of the experimental cantilever. The phase extremely decreases almost linear with the frequency (notice the phase change from $0^0$ to $-4200^0$): this is due to the phase change in the fixed cables, which deliver the electronic drive signal to the cantilever.
The red lines indicate the experimentally determined resonance frequencies, which are indicated at the top together with their corresponding Q-factors that describe the widths of the resonance peaks. The second panel from the top shows the spectrum of the numerical cantilever for comparison. The blue lines indicate the frequencies of this particular excitation scheme with $f_t = 2.50$ MHz (cantilever) and $f_s = 2.52$ MHz (sample), as well as the difference frequency $f_{diff} = 20$ kHz.\label{sfig3}}
\end{figure}

On the basis of the resonance frequencies $f_i^{num}$ of the numerical cantilever and the corresponding resonance frequencies $f_i^{exp}$ of the experimental cantilever, we defined a normalized, relative error $e_i$ for each resonance frequency:

\begin{equation}
e_i = \frac{|f_i^{num}-f_i^{exp}|}{f_i^{exp}}\label{s4-e1}
\end{equation}

We took into account the first 5 modes of the cantilever and used the average $e_i$ as a measure for the quality of our fit. We optimized the fit by varying the elasticity $E_t$ of the cantilever, the length $L$ of the cantilever, the tip mass $m_e$, the moment of inertia $I_e$ of the tip, and the density $\rho_s$ of the cantilever. As a best fit, with an average error of 1.397\%, the cantilever is described by following parameters: $E_t = 222$ GPa, $L = 207$ $\mu$m, $m_e = 5.76 \cdot 10^{-15}$ kg, $I_e = 3.51 \cdot 10^{-22}$ kg m$^2$, and $\rho_s = 3207$ kg m$^{-3}$. We did not fit the width and the thickness of the cantilever. Instead we have chosen them to be 20 $\mu$m and 2.7 $\mu$m, respectively, to set the spring constant of the numerical cantilever to 2.5 N/m such that it is comparable to the spring constant of $2.7 \pm 0.4$ N/m of the experimental cantilever.

The Q-factors that describe the widths of the resonance peaks, were chosen such that the widths of the resonance peaks match between the numerical and the experimental cantilever. If $Q_i^{exp}$ is the experimentally measured Q-factor of the resonance frequency $f_i^{exp}$, the corresponding numerical Q-factor $Q_i^{num}$ is related to $Q_i^{exp}$ by

\begin{equation}
Q_i^{num} = \frac{f_i^{num}}{f_i^{exp}} Q_i^{exp}\label{s4-e2}
\end{equation}

Fig. \ref{sfig3} shows the vibration spectrum of both the experimental cantilever and the numerical cantilever. The bottom two panels show the amplitude and the phase of the experimental cantilever. The phase extremely decreases almost linear with the frequency (notice the phase change from $0^0$ to $-4200^0$): this is due to the phase change in the fixed cables, which deliver the electronic drive signal to the cantilever.
The red lines indicate the experimentally determined resonance frequencies, which are indicated at the top together with their corresponding Q-factors that describe the widths of the resonance peaks. The second panel from the top shows the spectrum of the numerical cantilever. The blue lines indicate the frequencies of this particular excitation scheme with $f_t = 2.50$ MHz (cantilever) and $f_s = 2.52$ MHz (sample), as well as the difference frequency $f_{diff} = 20$ kHz.

\subsection{7. Complete Overview of the Results of the Numerical Calculations\label{SSec6}}

In the main text, we describe the results of three different schemes for the ultrasonic excitations: {\it off-off resonance}, in which the ultrasonic excitations are chosen halfway between the $3^{rd}$ and $4^{th}$ resonance of the cantilever; {\it off-on resonance}, in which the ultrasonic excitation frequencies are on the $4^{th}$ resonance of the cantilever; {\it experimental excitation}, in which the ultrasonic excitation frequencies are equally far away from the nearest resonance frequency as in the experiment. In this section, we present the full numerical results of both the {\it off-off resonance} and the {\it off-on resonance} scheme.

\begin{figure}[!ht]
\begin{center}
\includegraphics[width=125mm]{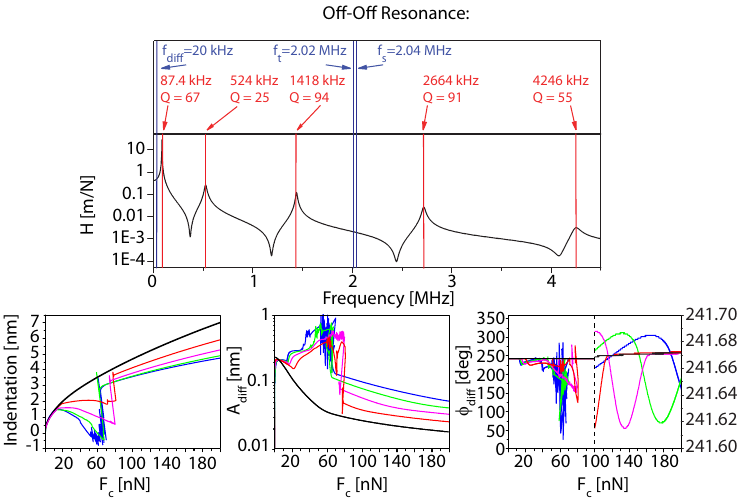}
\end{center}
\caption{The top panel shows the vibration spectrum in the {\it off-off resonance} scheme: blue lines indicate the excitation frequencies, whereas red lines indicate the resonance frequencies of the cantilever. The bottom three panels show the numerical results of the {\it off-off resonance} scheme for different sample elasticities: 2 GPa (black), 3 GPa (red), 4 GPa (magenta), 5 GPa (green), and 6 GPa (blue). The panels display from left to right the indentation ($\sim$ inverted height), the amplitude $A_{diff}$, and the phase $\phi_{diff}$. We observe an instability in the cantilever's motion while indenting into the sample (see jumps at $\sim 70$ nN). Note that, after the instabilities, we receive smooth motions of the cantilever and the phase differences approach values that correspond to the numerical error of the lock-in.\label{sfig4B}}
\end{figure}

Figure \ref{sfig4B} shows the results for the {\it off-off resonance} excitation scheme. The top panel shows the vibration spectrum: blue lines indicate the excitation frequencies and red lines the resonance frequencies. The bottom three panels show the numerical results for different sample elasticities: 2 GPa (black), 3 GPa (red), 4 GPa (magenta), 5 GPa (green), and 6 GPa (blue). The panels depict from left to right the indentation ($\sim$ inverted height), the amplitude $A_{diff}$, and the phase $\phi_{diff}$. The cantilever's motion is unstable for some contact forces while indenting in the sample (see jumps at $\sim 70$ nN). Note that after the instability the motion is stable again, $A_{diff}$ is smooth, we receive smooth motions of the cantilever, and the phase differences approach values that correspond to the numerical error of the lock-in.

\begin{figure}[!ht]
\begin{center}
\includegraphics[width=125mm]{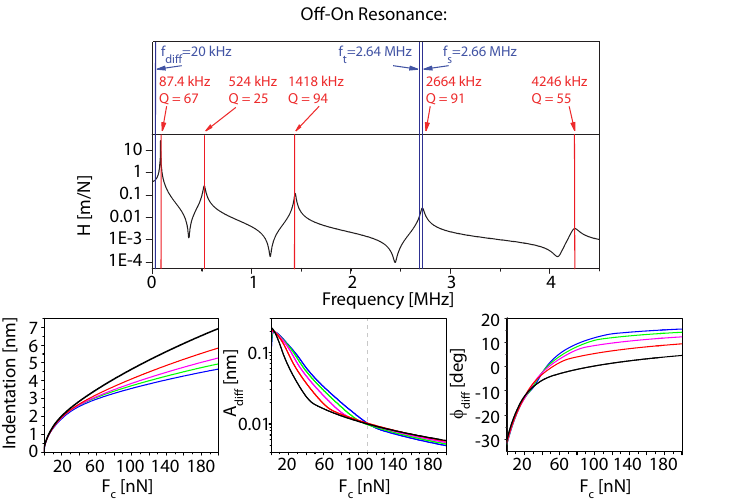}
\end{center}
\caption{The top panel shows the vibration spectrum in the {\it off-on resonance} scheme: blue lines indicate the excitation frequencies, whereas red lines indicate the resonance frequencies of the cantilever. The bottom three panels show the numerical results of the {\it off-on resonance} scheme for different sample elasticities: 2 GPa (black), 3 GPa (red), 4 GPa (magenta), 5 GPa (green), and 6 GPa (blue). The panels display from left to right the indentation ($\sim$ inverted height), the amplitude $A_{diff}$, and the phase $\phi_{diff}$.\label{sfig4A}}
\end{figure}

Figure \ref{sfig4A} shows the results for the {\it off-on resonance} excitation scheme. The top panel shows the vibration spectrum: blue lines indicate the excitation frequencies and red lines the resonance frequencies. The bottom three panels show the numerical results for different sample elasticities: 2 GPa (black), 3 GPa (red), 4 GPa (magenta), 5 GPa (green), and 6 GPa (blue). The panels depict from left to right the indentation ($\sim$ inverted height), the amplitude $A_{diff}$, and the phase $\phi_{diff}$.

This time, we do not observe any instabilities, as the value of the transfer function of the cantilever decreases with the shifting of the resonance frequencies towards higher frequencies. This is also the reason, why we do not see an instability in the results of the {\it experimental excitation} scheme, which is presented in the main text.
Let us, in the following, have a closer look to the implications on the contrasts, if applying a the specific excitation scheme.

We start with the height contrast. In both the {\it off-off resonance} and the {\it off-on resonance} excitation scheme, we observe that a softer sample (2 GPa) leads to a deeper indentation at a given contact force. Since we consider measurements that are performed with the feedback operating in contact mode, the contact force is held constant and a variation in elasticity results in different indentations, which translates into a measurable height signal: a harder material appears to be higher. This consideration holds for all excitation schemes including also the {\it experimental excitation} scheme, as discussed in the main text.

Considering the contrast in the amplitude $A_{diff}$ that results from parts of the sample with different elasticities, we observe opposite behavior between the {\it off-off resonance} scheme and the {\it off-on resonance} scheme. In the {\it off-off resonance} excitation scheme, we see that a hard surface leads to a higher amplitude $A_{diff}$ than a soft surface. This additionally supports both the experimental results of Sect. 4 and the analytical result of Sect. 5. In contrast, in the {\it off-on resonance} excitation scheme, we observe that for large contact forces ($> 110$ nN, see Fig. \ref{sfig4A}), a soft surface generates a higher amplitude $A_{diff}$ than a hard one. We trust this result of our simulation at large contact forces, as the cantilever is completely vibrating in the Hertzian contact regime of the tip-sample interaction at these contact forces: the cantilever does not feel any attractive forces during is motion. This is not the case at lower forces ($< 110$ nN), where the contrast is inverted. Further evidence for a contrast inversion as a function of the applied contact force comes from the fact that the retract curves (not shown here) show exactly the same characteristics. The contrast inversion between the {\it off-on resonance} case and the {\it off-off resonance} case is caused by the frequency shift of the $4^{th}$ resonance frequency of the cantilever, which is explicitly excited in the {\it off-on resonance} excitation scheme. 

\subsection{8. Resonance Frequency of the ``Nanoparticle in Polymer'' system\label{SSec9}}

For the estimation of the resonance frequency of the system {\it gold nanoparticle in polymer}, we need the spring constant of the PVA as well as the PMMA layer that are above and below the nanoparticle, respectively. From the stress equations of the polymer layers (see Eq. \ref{s3-e1}), we find the following spring constants:

\begin{align}
k_{PVA} &= \frac{E_{PVA} \pi R^2}{d - R} \approx 8.7\,{\rm N/m,}\\
k_{PMMA} &= \frac{E_{PMMA} \pi R^2}{t - d - R} \approx 7.4\,{\rm N/m,}
\end{align}

\noindent
for which we used the same physical values as in Supplementary Note 6: $R = 10$ nm, $d = 97$ nm, $t = 209$ nm, $E_{PVA} = E_{PMMA} = 2.4$ GPa.

Next we need the total mass $M$ of the spherical nanoparticle. As the mass density, $\rho_{Au}$, of gold is 19300 kg/m$^3$, we find:

\begin{equation}
M = \tfrac{4}{3}\pi R^3 \rho_{Au} = 8.1 \cdot 10^{-20}\,{\rm kg}
\end{equation}

By assuming a simple harmonic oscillator, in which two springs are attached to a mass, we find an estimation for the resonance frequency:

\begin{equation}
\omega_0 = \sqrt{\frac{k_{PVA} + k_{PMMA}}{M}} = 1.4 \cdot 10^{10}\,{\rm rad/s.}
\end{equation}

This results in a resonance frequency of 2.2 GHz.

\section*{Supplementary References}

\end{document}